\documentclass[aps,prd,twocolumn,nofootinbib,superscriptaddress]{revtex4}
\usepackage{graphicx}
\usepackage{amsmath,slashed}
\usepackage{microtype}

\def\ltap{\ \raise.3ex\hbox{$<$\kern-.75em\lower1ex\hbox{$\sim$}}\ }
\def\gtap{\ \raise.3ex\hbox{$>$\kern-.75em\lower1ex\hbox{$\sim$}}\ }
\newcommand{\gsim}{\lower.7ex\hbox{$\;\stackrel{\textstyle>}{\sim}\;$}}
\newcommand{\lsim}{\lower.7ex\hbox{$\;\stackrel{\textstyle<}{\sim}\;$}}
\def\OO{{\cal O}}

\newcommand{\GeV}{\,\mathrm{GeV}}

\def\unit{\relax{\rm 1\kern-.26em I}}
\newcommand{\half}{{\frac{1}{2}  }}

\newcommand{\MET}{\slashed{E}_T}
\newcommand{\go}{\tilde{g}}

\begin{document}
\pagestyle{plain}

\title{High Multiplicity Searches at the LHC Using Jet Masses}

\author{Anson Hook}
\affiliation{SLAC, Stanford University, Menlo Park, CA 94025}
\affiliation{Applied Physics Department, Stanford University, Stanford, CA 94305 }

\author{ Eder Izaguirre}
\affiliation{SLAC, Stanford University, Menlo Park, CA 94025}
\affiliation{Physics Department, Stanford University, Stanford, CA 94305 }

\author{ Mariangela Lisanti}
\affiliation{Princeton Center for Theoretical Science, Princeton University, Princeton, NJ, 08544}

\author{Jay G. Wacker}
\affiliation{SLAC, Stanford University, Menlo Park, CA 94025}
\affiliation{Stanford Institute for Theoretical Physics, Stanford University, Stanford, CA 94305 }

\date{\today}

\begin{abstract}
This article introduces a new class of searches for physics beyond the Standard Model that improves the sensitivity to signals with high jet multiplicity. The proposed searches gain access to high multiplicity signals by reclustering events into large-radius, or ``fat," jets and by requiring that each event has multiple massive jets.  This technique is applied to supersymmetric scenarios in which gluinos are pair-produced and then subsequently decay to final states with either moderate quantities of missing energy or final states without missing energy.   In each of these scenarios, the use of jet mass improves the estimated reach in gluino mass by 20\% to 50\% over current LHC searches.
\end{abstract}

 \maketitle

\section{Introduction}

Many beyond the Standard Model theories exhibit the striking feature of predicting high-multiplicity final states with ten or more final state colored partons at the LHC.  In supersymmetric versions of the Standard Model \cite{Dimopoulos:1981zb,Nilles:1983ge,Martin:1997ns}, these final states typically arise from new colored particles that cascade decay through intermediate states such as neutralinos or charginos, decays into several top quarks, or from the Lightest Supersymmetric Particle (LSP) decaying via baryonic R-parity violating couplings.  Other theories, such as those with strongly coupled electroweak symmetry breaking \cite{Evans:2009ga,Hill:1991at, Agashe:2003zs,Agashe:2006hk}, also give rise to final states with many jets, frequently from multiple top quarks through processes  such as $\rho_T \rightarrow 2\pi_T \rightarrow 2(t\bar{t})$ or $\omega_T \rightarrow 3 \pi_T\rightarrow 3(t\bar{t})$.   

Collider searches for high-multiplicity final states are challenging for two main reasons.  
The first  challenge is that the jets tend to be relatively soft.  The typical jet energy in top decays is around \mbox{60 GeV}, and after projecting onto the transverse direction, falls near the LHC jet energy threshold of \mbox{50 GeV}. 
An additional challenge is that many such signals typically have suppressed missing energy, making the events more difficult to separate from QCD and electroweak backgrounds.  For cascade decays, the presence of additional particles in the final state converts missing energy to visible energy.  In top-rich final states, not much phase space is available for the decay, thereby resulting in a low-momentum LSP.  

Studies of Simplified Models~\cite{Alves:2011wf} that approximate high-multiplicity supersymmetric topologies find that they are challenging to discover with standard searches that cut on the missing ($\MET$), visible ($H_T$), or total energy ($S_T$)~\cite{Alves:2011sq, Lisanti:2011tm, Essig:2011qg, Brust:2011tb,Papucci:2011wy,Kats:2011qh,Kane:2011zd, Bramante:2011xd}.  Dedicated searches for high-multiplicity final states exist at the LHC, including searches for top-dominated decays~\cite{ATLAS-CONF-2011-098, ATLAS-CONF-2011-130, CMS-PAS-SUS-11-006, CMS-PAS-SUS-11-010}, six or more jets and missing energy~\cite{Aad:2011qa}, and black hole resonances~\cite{CMS-PAS-EXO-11-071, ATLAS-CONF-2011-068}.  These searches typically require fairly significant cuts on missing energy and/or on $S_T$ to reduce QCD and top background contributions.  For multi-top topologies, jet substructure is a useful tool for reconstructing the top masses \cite{Gregoire:2008mr,Gregoire:2011ka,Berger:2011af}; these are typically low efficiency searches, however, and more inclusive searches should increase sensitivity in the discovery phase of the LHC.

This article proposes a new framework for discovering high-multiplicity final states that makes use of modern jet algorithms.  The proposal is to search for events with multiple ``fat'' jets, and to use information about the mass of these jets to discriminate against background.  In this work, a ``fat'' jet is defined using an anti-$k_T$ algorithm with $R=1.2$~\cite{Cacciari:2008gp}, though other definitions such as $R=1.0$ or Cambridge-Aachen jet clustering~\cite{Dokshitzer:1997in, Wobisch:1998wt} may work equally as well.  The jet mass variable we use is
\begin{eqnarray}
M_J = \sum_{i=1}^{n_J} m_{j_i},
\label{eq: totalmass}
\end{eqnarray}
where $m_{j_i}$ is the mass of the $i^{\text{th}}$ jet and $n_J$ is the total number of jets in the event. We explain its effectiveness over standard handles like $H_T$ as a background discriminator in Sect.~\ref{sec: Observables}.  In Sect.~\ref{sec: limits}, we discuss an implementation of a jet mass search and show how it can dramatically improve limits on high-multiplicity signals.  We conclude in Sect.~\ref{sec: discussions} with a discussion on generalizations of the searches presented here, as well as suggestions for data-driven background estimates. 

\section{Jet Mass as an Observable}
\label{sec: Observables}
Jet masses have historically been difficult observables at hadron colliders because pile-up and underlying event contribute to the jet mass as $R^3$ or $R^4$.  However, using jet-grooming techniques such as filtering~\cite{Butterworth:2008iy}, pruning~\cite{Ellis:2009me}, or trimming~\cite{Krohn:2009th}, the underlying event and pile up contributions can be removed.  The resulting jet is an accurate measurement of the underlying partonic event~\cite{Miller:2011yd,Miller:2011qg}.  Of these three methods, filtering is the least optimal for high multiplicity signals because it requires a fixed number of subjets to be identified in advance, whereas the signals studied in this article do not have a definite number of subjets per jet. 

The jet-grooming techniques listed above are designed to look for boosted hadronic resonances appearing under a continuum background.  The kinematics considered in this article typically result from particles decaying at rest and hence, the reconstructed jets do not group the underlying partons together in any  manner that represents the underlying decay kinematics.  As a result, the jet masses do not correspond to a parent particle's mass.  While jet-grooming with a variable number of subjets may be useful or beneficial, it is not as necessary and the details are not as important.  For the remaining portion of the article, no jet-grooming is used, but it should be understood that jet-grooming can be applied so long as the algorithm allows the number of subjets per fat jet to vary on a jet-by-jet basis.  In addition,  it may be possible to combine Qjets with jet pruning to even better improve sensitivity over background~\cite{Ellis:2012sn}.
\begin{figure}[t] 
   \centering
   \includegraphics[width=0.40\textwidth,angle=0]{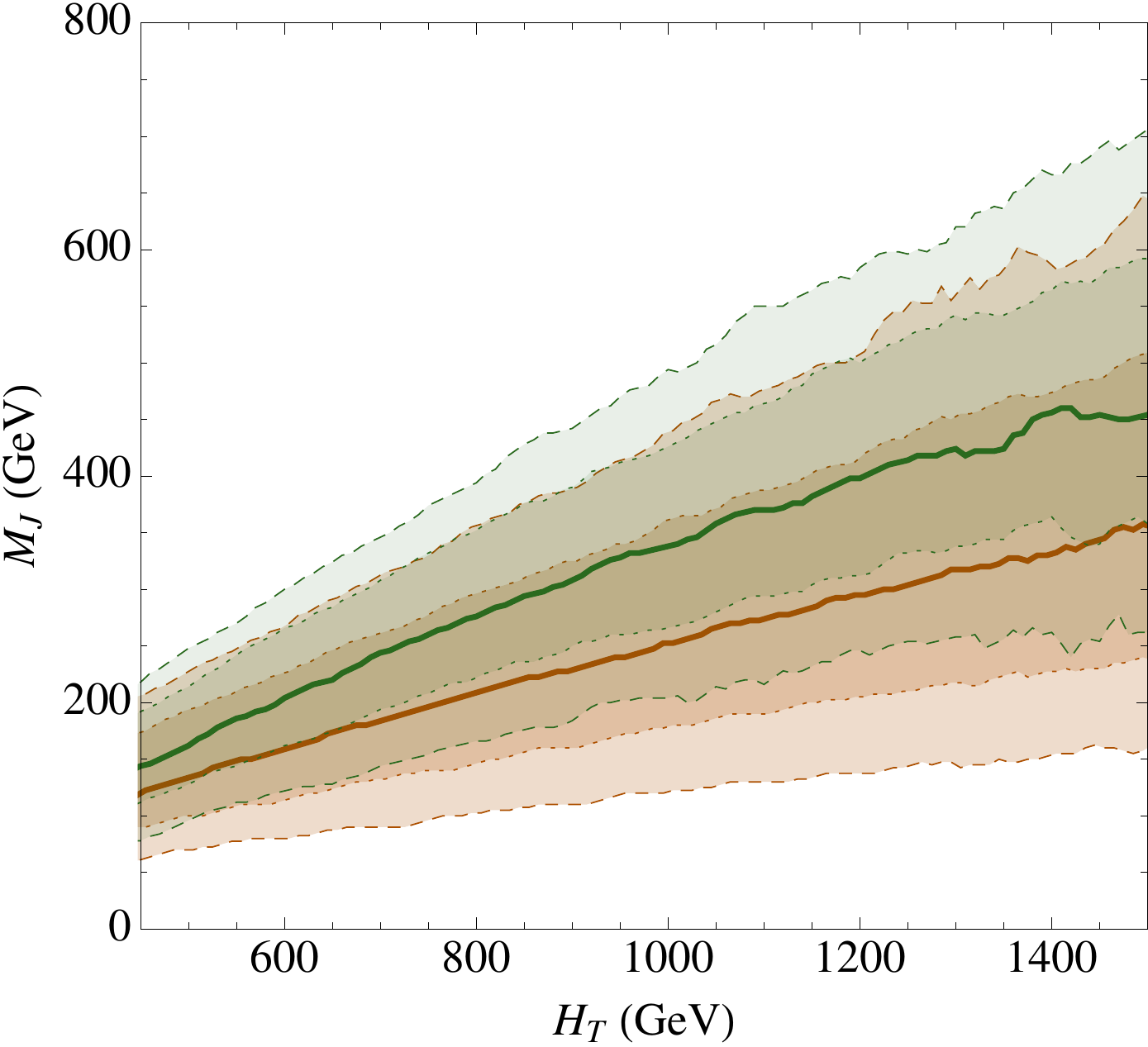} 
   \caption{A plot of $M_J$ versus $H_T$ after requiring $N_j\ge4$ ``fat'' jets with $p_T>120\GeV$ and $p_T>50\GeV$ on the leading and sub-leading jets, respectively.  QCD (orange) and top (green) events are shown where the median value for a given $H_T$ is shown in a solid line and the 68\% and 95\% inclusion bands are shown in the dotted and dashed lines, respectively.  The higher values of $M_J$ for top events arise from the top mass.  Signals with heavier parent particles than the top give even larger $M_J$.  }
   \label{fig: jet mass}
\end{figure}

\begin{figure*}[t] 
   \centering
   \includegraphics[width=0.75\textwidth,angle=0]{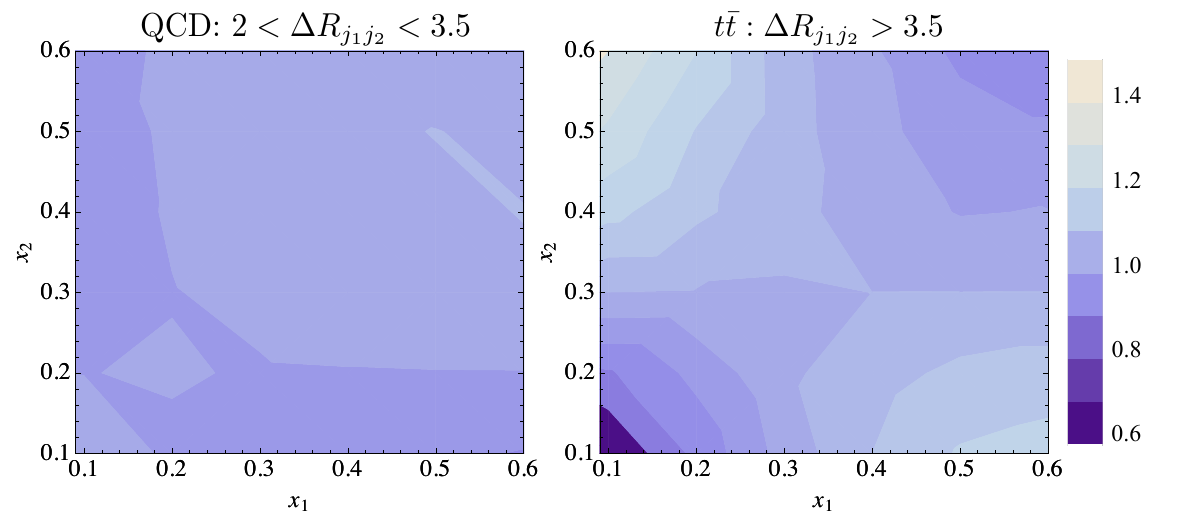} 
   \caption{The jet mass correlation, $H(x_1,x_2)$, for the hardest two jets in QCD (left) and $t\bar{t}$ (right) events that are clustered into three fat jets, where $x_i=m_{j_i}/p_{T,i}$.  A mild positive correlation is shown for QCD events, while a sizable anti-correlation is shown for  $t\bar{t}$ events.}
   \label{fig: ttbar}
\end{figure*}

When a jet is formed via a parton shower, its mean squared invariant mass is $\langle m_{j_i}^2 \rangle \propto \alpha_s p_{T,i}^2R^2$, where $\alpha_s$ is the strong coupling constant, $p_{T,i}$ is the transverse momentum of the jet, and $R$ is its radius~\cite{Salam:2009jx, Ellis:2007ib}.  
When a jet is formed from independent partons through multi-body decays of heavy particles, however, the typical jet mass is larger.  In high-multiplicity signal events, there is not enough solid angle for the partons to be well-separated and therefore multiple partons are clustered together.
As a result, partons will lie close to each other and may be clustered together into the same jet.  For these jets, the mean squared invariant mass is $\langle m_{j_i}^2 \rangle \propto p_{T,i}^2R^2$, where one does not pay the factor of $\alpha_s$.   

The visible energy in the event, $H_T$, can be related to the total jet mass $M_J$.  In particular, 
\begin{eqnarray}
\nonumber
H_{T} &=& \sum_{i=1}^{n_J} (p_{T,i}^2 + m_{j_i}^2)^{\half}\\ 
& \propto& \sum_{i=1}^{n_J} \sqrt{\langle m_{j_i}^2 \rangle((\kappa R)^{-2} +1)} \simeq M_J \frac{\sqrt{ 1+ (\kappa R)^2}}{ \kappa R},
\label{eq: HT}
\end{eqnarray}
where $\kappa = \sqrt{\alpha_s}$ for  jets whose mass is generated by the parton shower and 1 for jets whose mass arises from multiple partons being grouped together.  
Eq.~\ref{eq: HT} is the main reason why $M_J$ is a more effective discriminator than $H_T$ for high-multiplicity signals.  For high-multiplicity signals, the jet masses do not usually result from parton showering ($\kappa = 1$), while for the QCD and $V+\text{jets}$ backgrounds (when $V$ decays into missing energy) they do ($\kappa = \sqrt{\alpha_s}$).  For signal and background events with similar $H_T$, the value of $M_J$ for the background will always be lower than that for the signal.  As a result, the signal distribution always has a longer tail of high-jet mass than the background, even if its $H_T$ distributions are similar.   The correlation between $M_J$ and $H_T$ is shown in Fig.~\ref{fig: jet mass} for QCD and top events.  Top events typically have higher values of $M_J$ for a fixed $H_T$, with a total jet mass that asymptotes to $2 m_t$.  Signal events have even larger values of $M_J$ than top events and asymptote to higher values.

The argument that $M_J$ is preferable to $H_T$ relies on two assumptions.   The first is that the signal has a larger $M_J$ than top events, which requires that the signal is at least as jet-rich as top events and has higher typical visible energies than top events.  This first assumption is true in many signals of beyond the Standard Model physics.  
  
 The second assumption implicit in Eq.~\ref{eq: HT} is that jet masses are uncorrelated with each other.  Specifically, if one jet mass is anomalously large, then the probability that the second jet has a larger mass is not more significant than would be expected.  
 One measure of the correlation between the masses of two jets is the following quantity 
\begin{equation}
H(x_1,x_2) = \frac{h(x_1,x_2) \int h(x_1, x_2) dx_1 dx_2 }{\int h(x_1,x_2) dx_1  \int h(x_1,x_2) dx_2 },     
\end{equation}
where $h(x_1,x_2)$ is the two-dimensional distribution of $x_1$ and $x_2$, and $x_i=m_{j_i}/p_{T,i}$.  For jets with uncorrelated jet masses, $H=1$.  

The correlation between QCD jet masses was measured using a large sample of events generated with {\tt MadGraph4.4 + PYTHIA6.4} using parton shower matrix element matching.  Specifically, {\tt MadGraph} ~\cite{Alwall:2007st} was used to generate the following processes
\begin{eqnarray}
\nonumber
pp\rightarrow n_j&\quad & 2\le n_j \le 4^+,
\end{eqnarray}
where the four-parton multiplicity is an inclusive jet sample.
Here, $j$ refers to light flavor quarks and gluons, only. The leading parton was divided into five separate bins with
\begin{eqnarray}
\nonumber
p_T =\{\text{50--100}, \text{100--200},\text{200--300},\text{300--400}, >400\}  \text{ GeV}.
\end{eqnarray}
The $n_j=2,3,4$ samples each had 1M, 1M, and 0.5M events per $p_T$ sample, respectively.  Each of these events was parton showered and hadronized in {\tt PYTHIA}~\cite{Sjostrand:2006za} 200 times resulting in 2.5B total events, which were reconstructed with {\tt PGS5} using the anti-$k_T$ jet algorithm with R=1.2~\cite{PGS}. 
 Events with exactly three jets with $p_T > 100 \GeV$ and with a veto on the fourth jet \mbox{($p_{T j_4} < 50\GeV$)} were then used to calculate $H(x_1,x_2)$.  For  events with $\Delta R_{j_1 j_2} > 3.5$, the jet mass correlation is 
\begin{eqnarray}
|H^{\Delta R_{j_1 j_2} > 3.5}(x_1,x_2)-1| \le 0.05  \pm 0.05 \text{(stat)}.
\end{eqnarray}
Events with $2.0 < \Delta R_{j_1 j_2} < 3.5$ have
\begin{eqnarray}
|H^{2.0 < \Delta R_{j_1 j_2} < 3.5}(x_1,x_2)-1| \le 0.10 \pm 0.05\text{(stat)}
\end{eqnarray}
and exhibit a small positive correlation, as shown in the left panel of  Fig.~\ref{fig: ttbar}.  This correlation is small and we conclude that that there is no evidence for strong correlations amongst jet masses in events where jets arise from parton showering. 

Fig.~\ref{fig: ttbar} also plots $H(x_1, x_2)$ for the $t\bar{t}$ background with a $\Delta R_{j_1j_2} \ge 3.5$ requirement between the two leading jets to isolate the region with the least amount of correlations.  Both $W$ bosons from the top decays were forced to decay into hadrons.  For this example, the jet masses should be correlated, because they share a massive progenitor.  Indeed, as the figure shows, $H(x_1,x_2)$ deviates significantly from unity, and shows an anti-correlation between the two jet masses.  This can be understood as follows.  Most $t\bar{t}$ events are produced near threshold without significant additional radiation.  After requiring that three fat jets be identified, the six final state partons are grouped into the three jets.  Because there is a fixed number of final state partons arising from the decay of the top quark, if one jet acquires multiple partons, then it will reduce the typical number of partons in the second leading jet.  Thus, it is a zero-sum game.  

Signal events where the jets arise from the decay of massive colored particles (i.e., gluinos) should also have anti-correlated jet masses, just like $t\bar{t}$.  Because $M_J$ involves a sum over masses, this anti-correlation is not significant, and does not contribute large corrections to Eq.~\ref{eq: HT}.  The jet masses for the QCD and $V+\text{jets}$ backgrounds are uncorrelated because they arise from radiative processes; in this case, the corrections to Eq.~\ref{eq: HT} are also negligible.

\section{Sensitivity of Jet Mass Searches}
\label{sec: limits}

\begin{figure}[tb] 
   \centering
   \includegraphics[width=0.40\textwidth,angle=0]{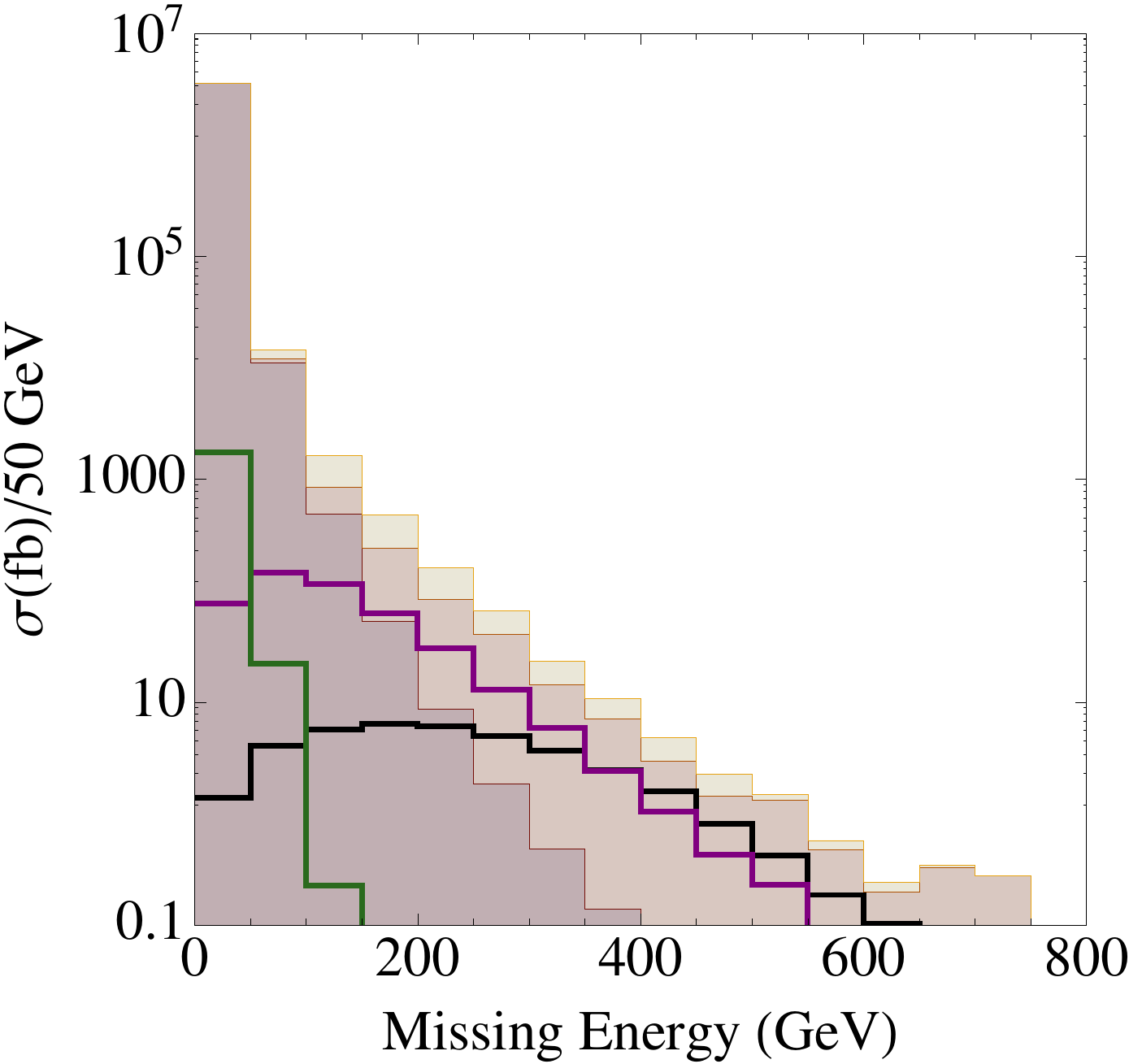} 
   \caption{Missing energy distributions for signal and background after requiring four or more fat jets.  Stacked histograms show the SM backgrounds, which include $t\bar{t}$ (light yellow), $V+nj$ (light red), and QCD (light purple).  The distributions for an 800 GeV gluino in the multi-top topology and a 600 GeV gluino in the 2-step cascade decay topology, both with a massless LSP, are shown in black and purple, respectively.  A 500 GeV gluino in the stealth SUSY topology is shown in green.}
   \label{fig: MET}
\end{figure}

To illustrate the improvement of $M_J$ searches over $H_T$ searches, we study two classes of signals, both arising from pair-produced gluinos $\tilde{g}$ that give rise to a large number of jets in the final state.  The first class consists of topologies with suppressed, but non-negligible missing energy, and the second class has hardly any missing energy.  Jet mass searches for these two classes will differ in whether a moderate missing energy requirement is necessary.  We consider each class separately in the following subsections.

\begin{figure*}[tb] 
   \centering
   \includegraphics[width=0.45\textwidth,angle=0]{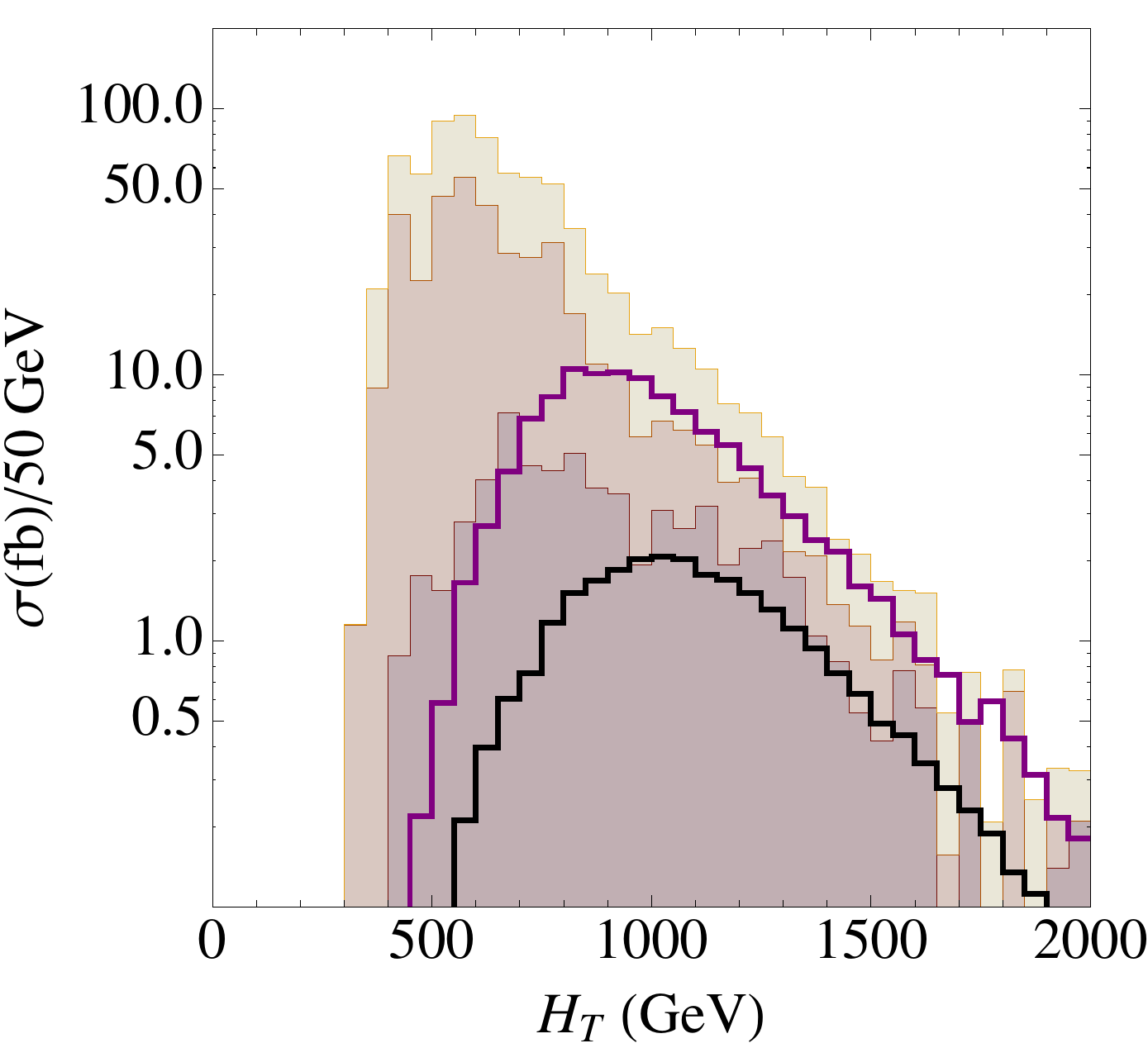} 
      \includegraphics[width=0.45\textwidth,angle=0]{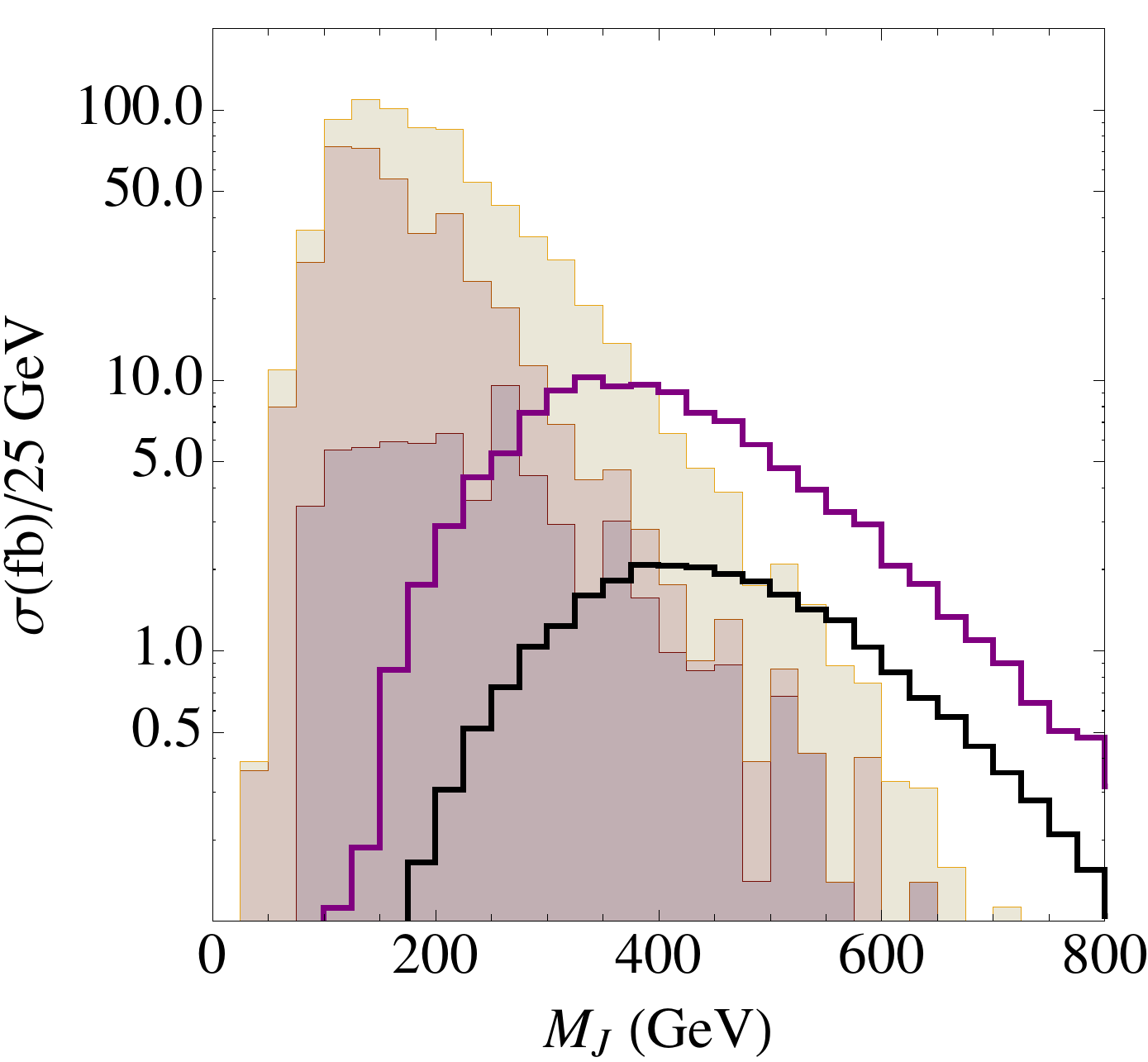} 
   \caption{(Left) $H_T$ distributions and (Right) $M_J$ distributions, after requiring four or more fat jets and $\MET> 150$ GeV.  Signal and background as in Fig.~\ref{fig: MET}. }
   \label{fig: HTMJ}
\end{figure*}
\subsection{Suppressed Missing Energy}

As examples of signals with suppressed missing energy, we consider a {\it multi-top} topology with
\begin{eqnarray}
\tilde{g} \rightarrow t\bar{t} + \chi
\end{eqnarray}
and a {\it 2-step cascade decay} topology with
\begin{eqnarray}
\go\rightarrow q q' \chi^{\pm}\rightarrow q q' W^{\pm}\chi'\rightarrow q q' W^{\pm}Z^{0} \chi,
\end{eqnarray}
where $\chi$ is the LSP, $\chi^{\pm}$ is a chargino, and $\chi'$ is a neutralino.  For the cascade topology, the chargino (neutralino) mass is halfway between that of the gluino (chargino) and LSP:
\begin{eqnarray}
m_{\chi^{\pm}}&=&m_{\chi}+(m_{\go}-m_{\chi})/2 \\ \nonumber
m_{\chi'}&=&m_{\chi}+(m_{\chi^{\pm}}-m_{\chi})/2.
\end{eqnarray}
This spectrum suppresses the missing energy significantly by reducing the available momentum to the LSP.  For this class of topologies, a modest cut on missing energy \mbox{($\MET > 100-150 \GeV$)} is useful, in addition to a cut on the jet masses.    

The samples of background and signal events used in the limit estimates were generated as follows. The parton-level signals for the multi-top and two-step cascade topologies were generated with {\tt MadGraph 4.4.44}~\cite{Alwall:2007st} in association with (up to) two jets
\begin{eqnarray}
pp\rightarrow \go\go + n_{j},
\end{eqnarray}
where $n_j=2$ for the highest multiplicity subprocess.  The importance of including additional radiation in signal processes has been documented in \cite{Alwall:2008ve, Alwall:2008va}.  To properly account for this initial-state radiation, we use the MLM parton shower/matrix element matching scheme~\cite{mlm} with a shower-$k_{\perp}$ scheme~\cite{Plehn:2005cq, Papaefstathiou:2009hp, Alwall:2009zu}.  The events are then showered and hadronized in {\tt Pythia 6.4}~\cite{Sjostrand:2006za}.  {\tt PGS 5}~\cite{PGS} is used as a detector mock-up and applies an anti-$k_T$ jet clustering algorithm with $R=1.2$~\cite{Cacciari:2008gp}.

The dominant Standard Model backgrounds include QCD, top production, and vector bosons plus jets.  The matched backgrounds are obtained for
\begin{eqnarray}
n_j, \qquad t\bar{t}+ n_t\qquad V+ n_v\qquad t+n_{t'}\qquad VV'+n_{V'}
\end{eqnarray} 
where $n_j=4$, $n_t= 2$, $n_v=3$, $n_{V'} = 2$, and $n_{t'} = 3$ are each the jet multiplicity of the highest-order process for each sample.  For $V+\text{jets}$, additional partons have been shown to be reasonably approximated by the parton shower \cite{Alwall:2007fs}.  The single-top and vector boson-pair production are subdominant and are thus not shown in the distributions in this paper, though they are included in the limit calculations.

Next-to-leading-order (NLO) corrections affect the normalization of both signal and background distributions.\footnote{With parton shower/matrix element matching, the shapes of differential distributions are accurately described by tree level predictions.}  The largest corrections are to the inclusive production cross section and can be absorbed into $K$-factors. The leading order cross sections of the signal are normalized to the NLO cross sections calculated in \texttt{Prospino 2.1} \cite{Beenakker:1996ch}. The leading order production cross sections for $t\bar{t}+\text{jets}$, $W^{\pm}+\text{jets}$, and $Z^0+\text{jets}$ are scaled to the NLO ones from \cite{Campbell:2010ff}. 

For the remainder of this article, the leading fat jet is required to have $p_{T\, j_1}>120$ GeV and the sub-leading fat jets have $p_{T}>50$ GeV. Fig.~\ref{fig: MET} shows the missing energy distributions for benchmark multi-top and cascade decay topologies with massless LSPs after requiring $N_j\ge4$.  Both these signals have events with missing energy above $\sim 100-200$ GeV, but not enough to effectively separate them from background.  

Figure~\ref{fig: HTMJ} shows the $H_T$ and $M_J$ distributions for these two benchmarks after a moderate missing energy cut of 150~GeV.  It is clear that the $M_J$ variable provides a far better discriminant against background than $H_T$, as expected from our discussion in the previous section.  By requiring several widely separated jets, QCD must produce these jets through an intrinsically $2\rightarrow 4$ process, as opposed to producing additional jets through the parton shower of a hard dijet event.  Requiring three or four fat jets plus a mild missing energy cut suffices in keeping QCD under control.  Electroweak vector bosons plus jets are subdominant backgrounds at low missing energy and are further reduced by the multiplicity requirement, especially at large jet mass.  

The dominant background comes from $t\bar{t}$ production, though the jet multiplicity and missing energy requirements help to keep it under control.  To pass these requirements, several of the jets must be grouped together to get sufficiently large jet mass and it is unusual to have two or more massive fat jets in top decays.  As discussed in Sect.~\ref{sec: Observables}, the jet masses from top quarks are more signal-like, in that they arise primarily from overlapping partons in the top decay.  Therefore, the total jet mass $M_J$ is not as suppressed as that for QCD.  However, the top quark events give rise to $M_J \lsim 2 m_t$, especially when at least one of the tops is forced to decay semileptonically by the missing energy requirement.  Therefore, a $M_J \gtrsim 400 \GeV$ is typically sufficient in removing the majority of the top background.
\begin{figure*}[tb] 
   \centering
   \includegraphics[width=0.45\textwidth,angle=0]{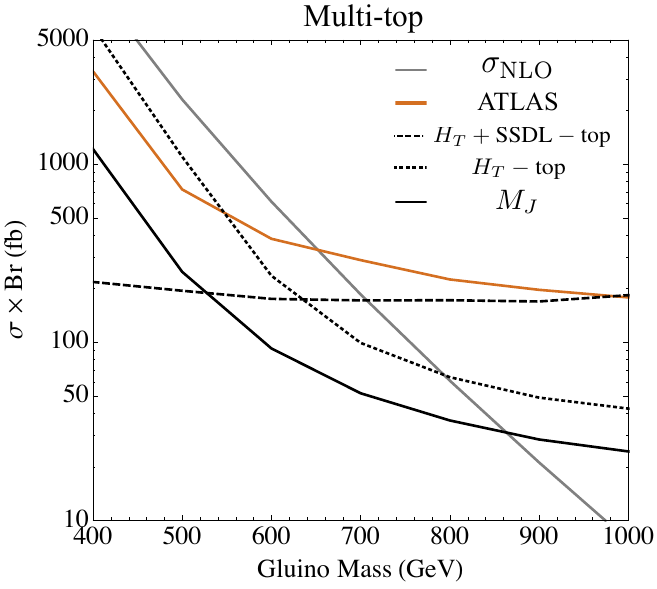} 
   \hspace{0.5in}
     \includegraphics[width=0.45\textwidth,angle=0]{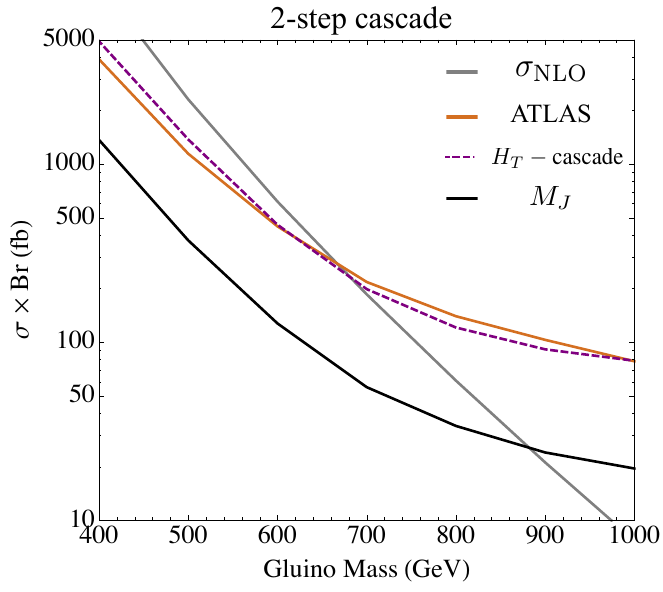} 
   \caption{Expected limit on $\sigma\times\text{Br}$ with \mbox{1.34 fb$^{-1}$} of integrated luminosity for the multi-top (left) and cascade decaying (right) topologies, assuming a massless LSP. The expected limit from an $M_J$ search after requiring events to have at least four fat jets with $\MET>150\GeV$ and $M_J>450\GeV$ is shown (solid black). The limit from ATLAS' high multiplicity search appears in orange \cite{Aad:2011qa}. The NLO production cross section for pair-produced gluinos is shown in grey. The expected sensitivity from optimal signal regions in $H_T$ are shown and are described in Tab.~\ref{table: stats}.}
   \label{fig: Limits}
\end{figure*}

Figure~\ref{fig: Limits} shows the expected 2$\sigma$ sensitivity to the multi-top and two-step cascade signals for a massless LSP, using \mbox{1.34 fb$^{-1}$} of integrated luminosity. The expected limits from optimal signal regions in $H_T$ are compared against the sensitivity of a $M_J$ search region. A 20\% systematic uncertainty on the backgrounds is assumed and is added in quadrature with the statistical error. The cuts that define each signal region are presented in Tab.~\ref{table: stats}.

The estimated limits from the current ATLAS large jet-multiplicity search~\cite{Aad:2011qa} are also shown in Fig.~\ref{fig: Limits} (orange lines).  The ATLAS search considers four signal regions with at least six, seven and eight jets. The stronger limit from the four signal regions is used for each gluino mass in Fig.~\ref{fig: Limits}.  In the ATLAS analysis, the jets are clustered using the anti-$k_T$ algorithm with R=0.4 and all pair combinations must satisfy $\Delta R > 0.6$.  An additional requirement that $\MET/\sqrt{H_T} > 3.5$ GeV$^{1/2}$ is enforced.  The reach of this search is comparable to that for the $H_T$ fat jet search, and is significantly weaker than that for the jet mass analysis. The event yields in the signal region from the $t\bar{t}$ Monte Carlo calculations in \cite{Aad:2011qa}  are in good agreement with  the $t\bar{t}$ generated in this study.

While we have only shown the estimated reach for the case of a massless LSP, we have found that the jet mass search also enhances the reach for arbitrary LSP masses.  However, different selection criteria are sometimes needed.  For instance, maintaining sensitivity for compressed spectra may require a weaker cut on $M_J$ and fewer massive jets. 

\begin{table}
\renewcommand{\arraystretch}{1.2}
  \begin{tabular}{  | l|  c  c  c  c  c  c c  |}
  \hline
    Search	 & $N_j$ &	R	 &   Leptons	& $N_{b}$ &  $\MET$  	 & $H_T$	&	$M_J$  	 \\
   	 &	 &	 &	 &  & [GeV]    	 &  [GeV]       &	[GeV]  	 \\
\hline
    ATLAS  	 &6-$8^{+}$&	0.4	 &  0	 &$0^+$ &	3.5 $\sqrt{H_T}$ &$\slashed{0}$ &	$\slashed{0}$ \\
    $H_T+$SSDL-top	&	$3^{+}$	 & 1.2	 & SSDL & $1^+$	 &	$\slashed{0}$	 & 300	 &$\slashed{0}$	 \\
     $H_T$-top	&	$4^{+}$	 & 1.2	 & $0^+$ & $1^+$	 &	250	 & 800	 &$\slashed{0}$	 \\
    $H_T$-cascade	&	$4^{+}$	&  	1.2	 & $0^+$ & $0^+$	 & 150 & 1000	 &$\slashed{0}$	 \\
    $M_J$ search &   $4^{+}$	 &     1.2    	 & $0^+$	 & $0^+$ &       150 &	$\slashed{0}$ & 450 \\
    \hline
  \end{tabular}
  \caption{The specifications of the searches used in Fig.~\ref{fig: Limits}.  A superscript ``+" indicates that the cut is inclusive.  ``SSDL'' denotes same-sign dileptons. }
  \label{table: stats}
\end{table}

\subsection{$\MET$-less Signals}

\begin{figure*}[tb] 
   \centering
   \includegraphics[width=0.45\textwidth,angle=0]{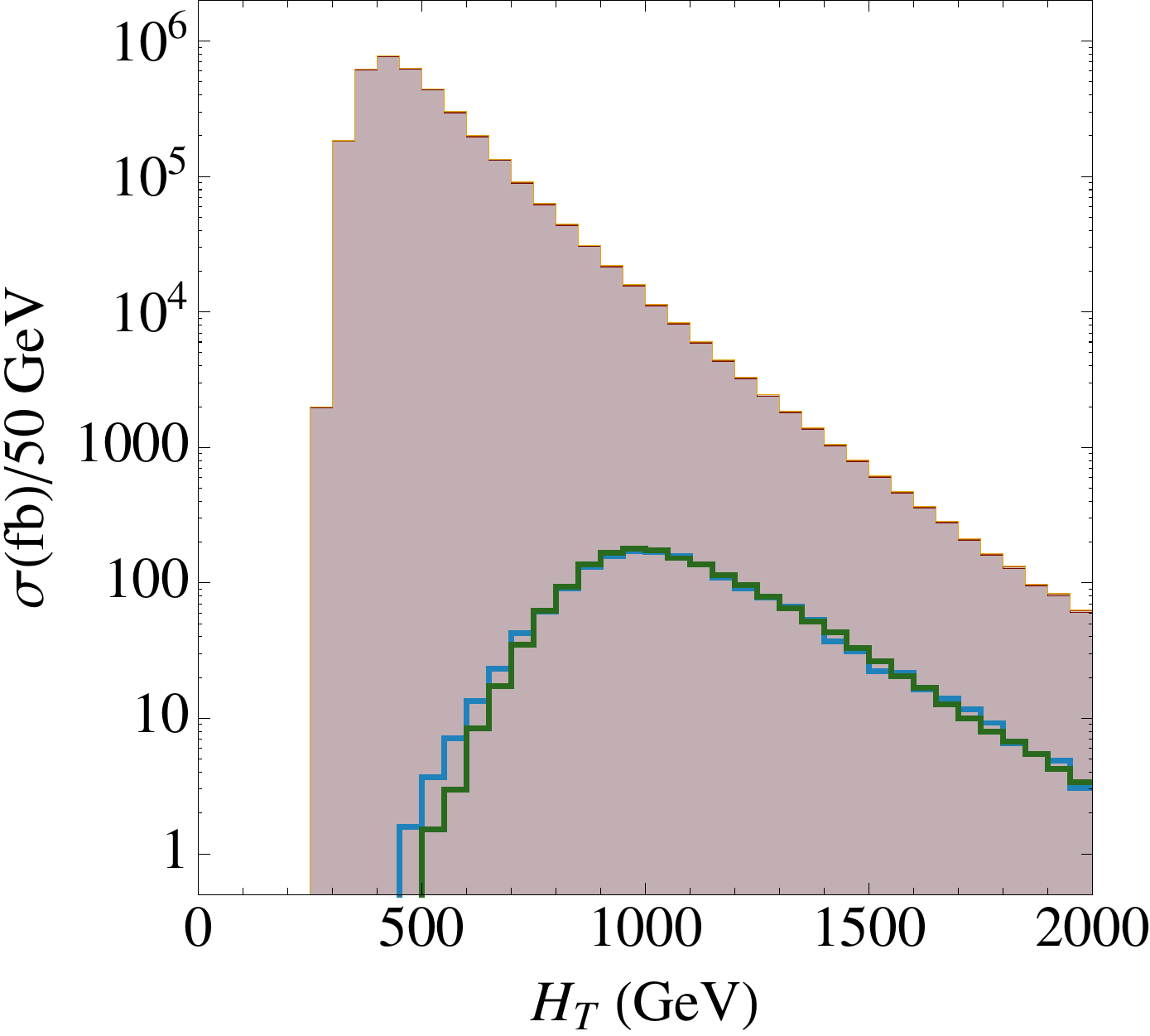}
   \hspace{0.5in} 
     \includegraphics[width=0.45\textwidth,angle=0]{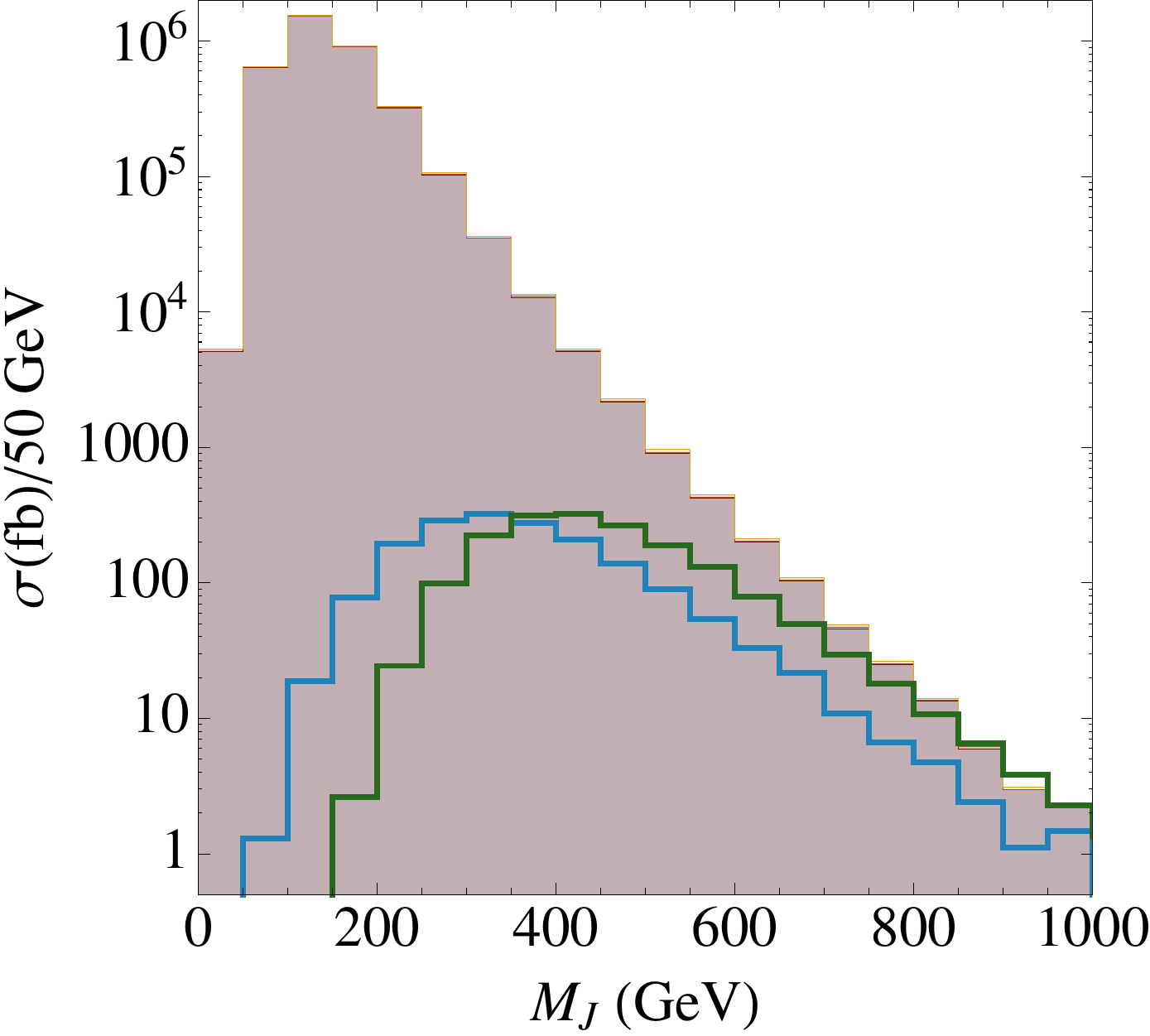} 
   \caption{(Left) $H_T$ and (right) $M_J$ distributions, after requiring four or more fat jets for backgrounds and 500 GeV gluinos decaying via RPV (blue) and stealth SUSY with $m_{\tilde{S}}=250\GeV$ and $m_{S}=220\GeV$ (green).  The backgrounds are shown stacked as in Fig.~\ref{fig: MET}, but are dominated by QCD.}
   \label{fig: StealthFig}
\end{figure*}

Next, we consider a class of topologies with hardly any missing energy ($\lesssim 100$ GeV).  Such models are challenging to separate from background because, without a missing energy requirement, the QCD background swamps the signal.  Black hole searches at the LHC do not have a missing energy requirement, but there, the signal dominates over background at $S_T$ greater than several TeV.  The SUSY topologies considered here have much lower $S_T$ and therefore would not be picked up by these searches.  

We will consider two examples of $\MET$-less signals here.  The first is a {\it stealth SUSY} topology \cite{Fan:2012jf,Fan:2011yu} with
\begin{eqnarray}
\go\rightarrow g \tilde{S} \rightarrow g \tilde{G} S \rightarrow g \tilde{G} gg,
\end{eqnarray}
where $\tilde{G}$ is the gravitino, and $S$ and $\tilde{S}$ are the singlet and singlino. For concreteness, we choose a spectrum where the singlino mass is half the gluino mass, and the singlino and singlet masses are split by only 30 GeV.  
The second is an {\it RPV SUSY} topology with
\begin{eqnarray}
\go\rightarrow 3 q.
\end{eqnarray}
The stealth SUSY signal is generated using \texttt{MadGraph}, while the RPV decaying gluinos are generated directly in \texttt{Pythia 6.4}.

Figure~\ref{fig: MET} compares the missing energy distributions of these signals with background.  Stealth SUSY does not have large intrinsic missing energy.  The RPV topology has no intrinsic missing energy and is therefore the more challenging of the two.  A missing energy cut of 150 GeV would eliminate both of these signals.  The standard ATLAS and CMS searches are applicable to stealth SUSY, especially the $\alpha_T$ search where no $\MET$ cut is used~\cite{Chatrchyan:2011zy}. However, they are sub-optimal given either the high $\MET$ requirements or in the $\alpha_T$ search, the similarity of the shape of signal and backgrounds. Currently, CMS has a dedicated search for RPV gluinos.  Instead of relying on missing energy, it searches for three-jet resonances in events with high jet multiplicity and large $H_T$.  The \mbox{35 pb$^{-1}$} analysis excludes gluino masses in the range from 200--280 GeV~\cite{Chatrchyan:2011cj}.

Fig.~\ref{fig: StealthFig} shows the $H_T$ and $M_J$ distributions for the stealth SUSY and RPV topologies after requiring $N_j\ge4$ fat jets.  Notice that the electroweak and top backgrounds are not even visible on the plot because of the overwhelming dominance of QCD.  The ratio of signal to background looks bleak when using $H_T$, however, $M_J$ provides a good variable with which to cut down QCD.  Fig. \ref{fig: StealthFig} shows that while a standard $H_T$ cut provides absolutely no sensitivity to stealth SUSY, a $M_J$ cut can increase the signal's significance by a factor of 50 and allow for bounds to be placed.  For the Stealth SUSY and RPV scenarios considered in this study, we find an expected limit on $m_{\go}$ of $\sim700\GeV$ and $400\GeV$, respectively, with 1 fb$^{-1}$ of luminosity. For RPV gluinos, using substructure on the leading and (possibly) sub-leading jets to reconstruct the gluino mass could complement the $M_J$ search \cite{Brooijmans:2010tn}.

\section{Discussion}
\label{sec: discussions} 

In this article, we show that a wide variety of high multiplicity signals for new physics models can be searched for by requiring several fat jets in an event, with large total jet mass.  A jet mass search is inclusive and increases the reach of the standard LHC searches to high multiplicity events.  Searches that explicitly require large numbers of standard-sized jets  suffer from the fact that, if a jet falls beneath the $p_T$ requirement of a hard jet, then it is not included in the event.  In essence, an $N$-jet search requires $\OO(N)$ cuts, reducing the overall efficiency.  Additionally, if partons accidentally fall near each other, they are clustered together and the jet multiplicity goes down.  The jet mass search proposed here is more inclusive for high multiplicity events because it does not explicitly place a requirement on the parton multiplicity and allows for more decay topologies to pass the event selection criteria. These searches are also inclusive in the number of $b$-jets and the number of leptons, which means that they are sensitive to the different exclusive signatures that multi-top events produce.  

The second benefit of using this more inclusive search is that it offers a better handle on backgrounds.   Computing high multiplicity Standard Model final states is challenging even at tree-level and NLO corrections remain beyond the reach of the current efforts.    Because jet mass is dominantly determined by the local parton shower evolution and approximately factorizes from the other jets in the event, this means that that the exclusive Standard Model calculation can be performed and that the jet mass function can be convolved with the process.  

The use of $M_J$ is quite robust at separating signal from background in a wide variety of contexts.  There are refinements to this search that could be useful.  $M_J$ is an inclusive variable but it could be augmented  by requiring that the jet mass arises from the existence of subjets rather than a diffuse source of energy that could come from the underlying event or pileup.  $N$-subjettiness \cite{Thaler:2010tr} is a natural variable to augment the searches.  For instance, requiring the existence of 6 or 8 subjets could potentially be powerful.  Another technique to identify subjets is ``prominence'' \cite{Jankowiak:2011qa} and could serve a similar function as $N$-subjettiness. 

The jet mass searches proposed in this article may be particularly well suited to run in the high luminosity  environment where 20 to 200 interactions per bunch crossing are typical.  In the high luminosity environment, the use of jet-grooming techniques will become critical to eliminate contamination of jet mass.  Alternatively, by using a track-mass rather than a calorimeter mass, the problem of pileup will be essentially eliminated.   The primary challenge is that tracking does not extend as far forward as calorimetry, but this direction offers promise and could potentially be used at the trigger level.

\section*{Acknowledgements}
The authors would like to thank Raffaele D'Agnolo, Daniele Alves, Matthew Reece, Gavin Salam, Philip Schuster, Matthew Strassler, and Natalia Toro for helpful discussions. We would like to thank  David W. Miller, Gavin Salam, and Ariel Schwartzman for helpful feedback on the draft.  ML acknowledges support from the Simons Postdoctoral Fellows Program.  AH, EI, and JGW are supported by the US DOE under contract no.~DE-AC02-76SF00515. EI is also supported by an LHCTI graduate fellowship under grant NSF-PHY-0969510. JGW is partially supported by the DOE's Outstanding Junior Investigator Award and the Sloan Fellowship.  ML was supported in part by the U.S. National Science Foundation, grant NSF-PHY-0705682.

\bibliography{MultiJetBibFile}

\end{document}